\documentclass[useAMS,usenatbib]{mn2e}
\usepackage{graphicx,amssymb}
\title[The central point source in G76.9+1.0]{The central point source in G76.9+1.0}
\author[Marthi et al.]{V.~R.~Marthi$^{1}$\thanks{\{vrmarthi, chengalur,
    ygupta\}@ncra.tifr.res.in}, J.~N.~Chengalur$^{1}$\footnotemark[1], 
  Y.~Gupta$^{1}$\footnotemark[1],
\newauthor G.~C.~Dewangan$^{2}$\thanks{\{gulabd,
  dipankar\}@iucaa.ernet.in} and D.~Bhattacharya$^{2}$\footnotemark[2]\\
$^{1}$National Centre for Radio Astrophysics, Tata Institute of
  Fundamental Research, Pune 411 007, India.\\
$^{2}$Inter-University Centre for Astronomy and Astrophysics, Pune 411 007, India.}
\begin{document}

\date{\today}

%\pagerange{\pageref{firstpage}--\pageref{lastpage}} \pubyear{2010}

\maketitle

\label{firstpage}

\begin{abstract}

  We describe the serendipitous discovery of a radio point source
  in a 618~MHz image of the supernova remnant(SNR) G76.9+1.0. The 
  SNR has a bipolar structure and the point source is located near 
  a faint bridge of emission joining the two lobes of emission.
  The point source was also detected in follow-up higher frequency 
  (1170~MHz) observations. The spectral index for the point source 
  obtained from the GMRT observations is $\alpha$ = -2.1$^{-0.45}_{+0.36}$. 
  The steep spectrum, as well as the location of the point source 
  near the centre of the SNR establish the fact that it is indeed 
  the pulsar J2022+3842 associated with this SNR. The weighted-mean 
  radio position of this point source is $\alpha =
  20^{h}22^{m}21.69^{s}\ \pm\ 0.11^{s},\ \delta =
  38\degr42\arcmin14.8\arcsec\ \pm\ 1.7\arcsec, J2000$. Consistent
  with this, subsequent analysis of archival {\it Chandra} X-ray data
  shows a point source coincident with the radio point source, as well
  as diffuse extended X-ray emission surrounding the unresolved
  source. However, no pulsed emission was detected despite deep
  searches at both 610~MHz and 1160~MHz although pulsed emission has been 
  seen at 2~GHz with the GBT. It appears that the most likely reason for not 
  detecting the pulsed signal at the GMRT is temporal broadening: 
  for the estimated DM towards this SNR, the pulse broadening time 
  could be as large as tens of milliseconds. The diffuse X-ray 
  emission is elongated along the same direction as the bipolar 
  structure seen in the radio. We interpret the radio lobes as 
  having been formed from an equatorial wind. Although direct detection
  of pulsed signal has not been possible, we show convincingly
  that sensitive, high-resolution, radio imaging at multiple frequencies
  is a useful method to search for pulsar candidates.

\end{abstract}


\section{introduction}
The source G76.9+1.0 was first identified by \citet*{wendker1991} in
a DRAO survey of the Cygnus-X region. Those authors tentatively 
identified it as a Galactic supernova remnant but, because of the
limited resolution of their survey, were unable to come to a firm
conclusion. From follow-up VLA multi-frequency observations
\citet*{landecker1993} showed it to have a two-lobed structure,
with the lobes themselves joined by a bridge of emission, i.e. morphologically
very similar to the pulsar wind nebula (PWNe) DA495 (see e.g. 
\citet{kothes2008} for a detailed discussion on DA495). From the 
observed rotation measure \citet{landecker1993} conclude that 
it is further than 7~kpc. The spectral index they measured for 
the source was  $\alpha = -0.62 \pm 0.04$ (where $S(\nu) 
\propto \nu^{\alpha}$). This is somewhat steep for a filled-center
remnant. Nonetheless, based on  its morphological similarity to DA495,
both \citet{landecker1993} and \citet{kothes2006} suggest that
G76.9+1.0 is also a PWNe. For G76.9+1.0 the integrated flux density at
the lowest 
available frequency (327~MHz) is substantially
discrepant: \cite{landecker1993} suggest that this might be  because
extended emission was resolved out in the 327~MHz image. Here we
present a GMRT 618~MHz image of this source. This image was obtained
as part of a test observation whose original aim was to test a scheme
for online system temperature correction of the visibility data. The
target was selected from \cite{green2009a, green2009b} for two
reasons: (i) it is located in the Galactic plane, where it would be
useful to test our experimental scheme, and (ii) the GMRT, being
sensitive to both large- and small-scale emission, could settle
questions on its unclear morphological type. 
The GMRT has a hybrid configuration which allows imaging of both 
the smooth extended emission as well as of fine scale structure 
with data taken in a single observing run. 
The observations and results presented in this paper 
however provide strong support for the interpretation that 
G76.9+1.0 is a pulsar wind nebula (PWNe).

The rest of this paper is organised as follows: in Section~\ref{gmrt610}, 
we describe our 610 MHz GMRT observations and results, which led to
the serendipitous detection of the radio point source. In
Section~\ref{psrch610} we describe a susequent search for pulsed
emission at 610~MHz, unaware that a search had already been carried
out \citep{zaven2010}. Since no pulsed emission was detected, we
followed up the 610 MHz observations with 1170 MHz observations, with
the aim of both mapping the point source as well as searching for
pulsed emission; the reasons behind this renewed search, the details
and results are described in Section~\ref{gmrt1160}. In order to
better understand the source, we also analysed archival \emph{Chandra}
data for this region. We present in Section~\ref{xray} the analysis
and modelling of \emph{Chandra} X-ray data. A discussion of the
results can be found in Section~\ref{summary}. 
\section{GMRT 610 MHz Observations}
\label{gmrt610}
\subsection{Observing and analysis details}
The supernova remnant G76.9+1.0 was observed on 24 Jun 2009 using
the GMRT hardware correlator. Flux and band pass calibration were
based on scans of the standard calibrators 3C286, 3C48 and 3C147; the
VLA calibrator 2052+365 was used for phase calibration. Although the
GMRT correlator produces 128 spectral channels across each of two
16-MHz wide sidebands, only the upper sideband data were used for
analysis because of problems with the data in the lower sideband. The
centre frequency of the resulting image is hence 618~MHz. Throughout
the observation, the  Automatic Level Control (ALC) was kept switched
off. Data reduction was  done using AIPS. Bad data were flagged, data
calibrated on a single channel and solutions applied to all channels
by bandpass calibration. The strong source 4C39.61 at the edge of the
field of view was first imaged and subtracted out. The rest of the
field was then imaged (and self-calibrated) using 37 facets. 
\begin{figure}
  \begin{center}
    \includegraphics[scale=0.45]{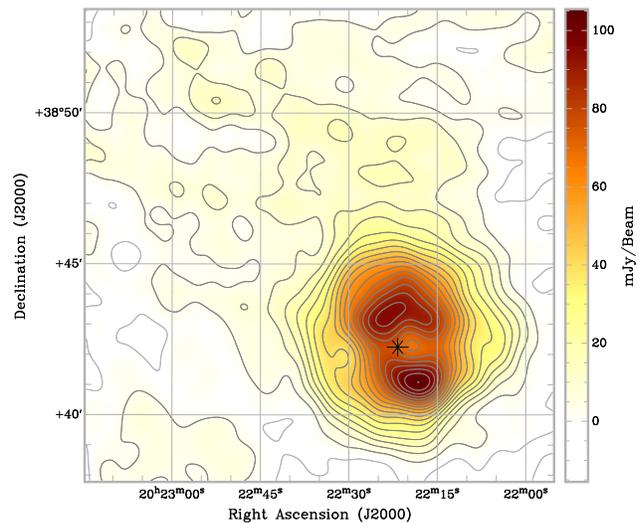}
    \caption{The 618 MHz continuum map of G76.9+1.0 obtained with the
    GMRT. The contours are at 5\% levels, the lowest contour is 20\%
    and the peak contour is 102 mJy/bm. The restoring beam used is
    54$\arcsec\ \times$  51$\arcsec$ so as to image the extended
    emission. The cross marks the position of a point source
    discovered in this data.}
    \label{610mhzmap}
  \end{center}
\end{figure}
\subsection{Results}
\label{ssec:610psrc}

\begin{figure}
  \begin{center}
    \includegraphics[angle=270,scale=0.33]{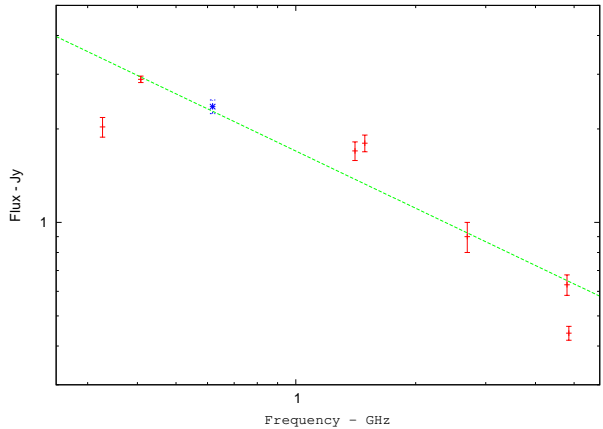}
    \caption{Spectrum of G76.9+1.0, with the GMRT 618-MHz
      measurement shown in blue. Integrated flux densities at 408 MHz,
      618 MHz, 2.695 GHz and 4.8 GHz give a power law fit with an index
      $\alpha=-0.61 \pm 0.03$ shown in green. }
    \label{spixfit}
  \end{center}
\end{figure}

\begin{figure}
  \begin{center}
    \includegraphics[scale=0.47]{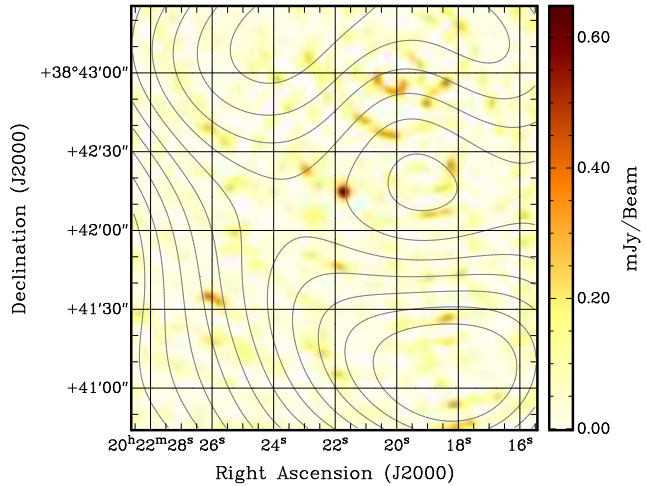}
    \caption{The point source seen at the position of the cross in
    Fig~\ref{610mhzmap} at 610~MHz. The contours of
    Fig~\ref{610mhzmap} are overlaid in this magnified view to highlight the
    location of the source in relation to the bridge of emission
    joining the two lobes and the central depression. The central
    point source has a peak flux density of 680 $\mu$Jy and the map RMS noise
    is 59 $\mu$Jy. The resolution of the map is 5$\arcsec\ \times$ 
    4.3$\arcsec$.}
    \label{610mhzpntsrc}
  \end{center}
\end{figure}

The low resolution 618~MHz continuum map is shown in Fig.~\ref{610mhzmap}.
The GMRT has a hybrid configuration \citep{swarup1991} with roughly half 
the antennas in a central region  approximately $1\times 1$~km$^2$ across. 
The remaining antennas are spread along three $\sim14$-km long arms, 
giving a maximum baseline length $\sim25$ km. This image is made using 
essentially only the data from antennas in the central square of the 
GMRT  (i.e. using a Gaussian UV taper with the 30$\%$ level at 2.5 k$\lambda$)
and has a resolution of $54\arcsec\ \times 51\arcsec$. The morphology matches 
well with the double lobed structure joined by a diffuse bridge of 
emission noted in earlier studies \citep{landecker1993, kothes2006}. 
The total integrated flux density from the remnant, uncorrected for
foreground contribution or absorption, is 2.36 Jy $\pm$ 0.28 Jy. The
RMS noise on the integrated flux is 45 mJy; however we have folded in
a 10\% error(0.236 Jy) to account for possible systematics(for this
observing mode with the GMRT, from similar observations in the past) in the
calibration. The GMRT 618-MHz flux measurement is plotted along with
other previous measurements (from \citealt{landecker1993}) in
Fig.~\ref{spixfit}. The earlier spectral index measurement obtained by
\citet{landecker1993} excluded certain measurements. Their reasons
included possible under-representation of large-scale emission due to
poor uv-coverage in the measurements they excluded for spectral
fitting. Including the new GMRT flux measurement, but following
\citet{landecker1993} in excluding the discrepant measurements at
327~MHz, 1408~MHz, 1490~MHz and 4850~MHz, the spectral index we
measure is  $\alpha = -0.61 \pm 0.03$ in excellent agreement with the
earlier measurement of $-0.62 \pm 0.04$. 
We used the same 618 MHz data to search for compact sources by making
an image after excluding all baselines shorter than 10k$\lambda$. A 
point source with a peak flux density of 680 $\mu$Jy $\pm$ 58 $\mu$Jy at
$\alpha: 20^{h}22^{m}21.71^{s}, \ \delta: 38\degr42\arcmin15.0\arcsec, J2000$
near the bridge of emission joining the two lobes is clearly
detected. This is shown in Fig.~\ref{610mhzpntsrc}, with the 618-MHz contours of
Fig.~\ref{610mhzmap} overlaid. Its location makes it a very promising
candidate for the central pulsar associated with this PWNe. The error
on the position is half of the synthesised beam - in this case
$\sim\ 2.5\arcsec$. The error on the position as returned by a
gaussian fit to the synthesised beam is minuscule, and as such makes
little sense as the positional uncertainty due to systematics is of
the order of the synthesised beam itself.  

\section{The search for a pulsar at 610 MHz}
\label{psrch610}
Subsequent to the discovery of the point source in the 610-MHz map,
and, as of yet unaware of the detection of pulsar J2022+3842, 
we decided to search for pulsed emission from the same object.
A pulsar search observation using the pulsar backend of the GMRT was carried out 
on 3 March 2010, with 13 antennas of the GMRT central square phased towards
the unresolved source discussed in Sec.~\ref{ssec:610psrc}. The total 
integration on the source was about 6 hours, with interspersed phasing 
scans of 5 minutes once every 45 minutes to an hour. The data were recorded 
with 256 channels for each of the two sidebands. 
In the image made from the simultaneously 
recorded inteferometric data (i.e. using the GMRT hardware correlator and
all available antennas) the point source was clearly detected.
The flux density measured at this epoch was  660 $\mu$Jy $\pm$ 80 $\mu$Jy,
consistent with the earlier measurement. For the assumed distance of
7~kpc the range of DMs given by the Galactic Free Electron Density
Model \citep{NE2001} is $\sim250\ pc\ \!cm^{-3}$. Since the true DM could differ
from this estimate by as much as a factor of 2, a search was made over
a DM range $1\ \!-\ \!500\ pc\ \!cm^{-3}$ in steps of $1\ pc\ \!cm^{-3}$. The pulsar search
pipeline at the GMRT uses a customised set of programs from both
SIGPROC and PRESTO \citep{presto}. The bad channels and sections of the pulsar search
data corresponding to the phasing scans were blanked out by applying
suitable time- and frequency-masks using the program
\emph{rfifind}. Subsequently the data were 
dedispersed using \emph{prespsubband} through the DM range mentioned
above. The dedispersed data  were Fourier-transformed using optimised
FFT routines. Finally an acceleration search using the program
\emph{accelsearch} \citep*{ransom2002} was done to search for
candidates. The limit of detectability attained as per calculations was
$\sim$100 $\mu$Jy per sideband. Although the phase-averaged flux
density of the unresolved source 
is significantly higher than the detectable limit, no pulsed emission was detected.

\section{GMRT L-band observations}
\label{gmrt1160}
The lack of pulsation was puzzling, since the location of this
source made it highly likely to be the central pulsar in this
supernova remnant. The fact that radio continuum emission was
detected, also ensures that this putative pulsar is beamed towards us.
Possible causes for this include (i)~the rotation and magnetic 
axes of the pulsar are aligned, (ii)~the pulse is significantly
broadened at low frequencies, and (iii)~the pulse period is near
20~ms where there is signficiant powerline-generated RFI. Since the
second of these seems apriori the most likely, a follow-up higher
frequency GMRT pulsar
search was made at 1170~MHz.   
\subsection{Observing and analysis details}
\label{gmrt1170obs}
G76.9+1.0 was observed at 1170~MHz with the GMRT on 16 June 2010.
Data were recorded simultaneously in the interferometric and 
phased array modes. All the available antennas were used to obtain
visibilities in the interferometric mode, whereas only the central 
square and the first arm antennas were added in phase to obtain 
the pulsar search data. The visibilities as well as the pulsar search
data were recorded separately as 256-channel, 
33-MHz baseband data with the GMRT Software Backend (GSB; \citealt{gsb}).
3C147 and 3C286 were used as primary calibrators for interferometric 
observations. 2052+365 was observed every 45 minutes for interferometric 
observations to enable phase  calibration and interpolation, which 
simultaneously also served as phasing scans for the pulsar search 
observation. The data reduction and analysis were very similar to 
that of the earlier 610-MHz observations. Regrettably, the upper half
of the 33-MHz bandwidth was not usable due to the presence of
intermittent RFI. Therefore, the centre frequency of the resulting
image is $\sim1160$~MHz. 
\subsection{Results}
\label{gmrt1180res}
\begin{figure}
  \begin{center}
    \includegraphics[scale=0.45]{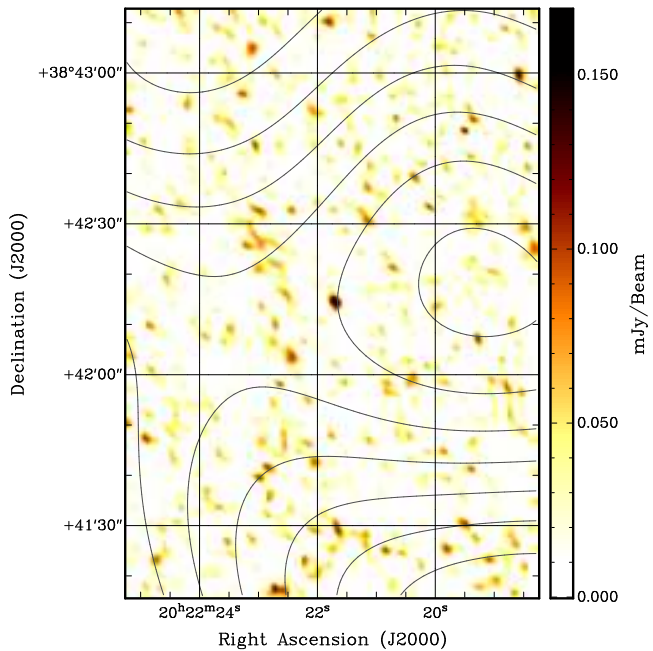}
    \caption{The point source seen at the same position at 1160 MHz
    with the GMRT. The beam in this full-resolution map is
    3.4$\arcsec\ \times$ 2.2$\arcsec$ and the map RMS noise is 22
    $\mu$Jy. The peak flux density of the unresolved source is $171$
    $\mu$Jy.} 
    \label{lbandpntsrc}
  \end{center}
\end{figure}

G76.9+1.0 is $\sim9\arcmin\times12\arcmin$ in size, and much of its flux is expected
to be resolved out at the GMRT at 1160~MHz. Hence no attempt was made
to measure its total flux at 1160~MHz. The full resolution image 
(Fig.~\ref{lbandpntsrc}) however shows a clear detection of the point
source seen earlier at 618~MHz. The position measured from the 
1160-MHz data ($\alpha:~20^{h}22^{m}21.67^{s},\ \delta:~38\degr42\arcmin14.5\arcsec,
J2000$)
is in excellent agreement with that measured at 618~MHz. Here again,
the positional uncertainty is $\sim\ 1.7\arcsec$. The measured
flux density at 1160~MHz is 171 $\mu$Jy $\pm$ 22  $\mu$Jy, giving a spectral 
index of $\alpha$ = -2.1$^{-0.45}_{+0.36}$. To cross-check the calibration,
we re-reduced archival 1465-MHz VLA data for this field. A comparison of
the flux densities measured by the VLA and the GMRT for the four brightest sources
is shown in Table~\ref{vlafluxcompare}. As can be seen, they 
are in good agreement. 

The pulsar search attained a detectability limit of 100 $\mu$Jy,
similar to the 610-MHz figure. Given that
the time-averaged flux density of the source at this same epoch is $\sim 170\ \mu$Jy,
one would have expected the pulsed emission to have been easily 
detected. However, no pulsed emission was detected despite searching
through a DM range of $1\ \!-\ \!500\ pc\ \!cm^{-3}$ (using the same procedure as
used earlier at 610~MHz). We discuss this further in
Sec.~\ref{ssec:nopulse}.

Given the lack of detection of pulsed emission, it is difficult to
assert with certainty that the point source that we see is indeed
the central pulsar in this SNR. X-ray observations could however
provide support to this interpretation. A search of the \emph{Chandra}
archival data revealed that this source had indeed been observed and
that data was available in the public domain. We discuss next archival 
{\it Chandra} X-ray data for this source. 

\begin{table}
\begin{tabular}{c|c|c}
\hline
RA,Dec 		&	 GMRT	 	&	 VLA		\\
(J2000)		&	 1160 MHz	&	 1465 MHz	\\ \hline
$20^h22^m20^s$ 	&                	&               	\\
38$\degr$33\arcmin59\arcsec 	&  2.0$ \pm $0.05 mJy 	& 4.98$ \pm $0.91 mJy	\\ \hline
$20^h23^m13^s$ 	&                	&                   	\\ 
38\degr38\arcmin23\arcsec       & 6.1$ \pm $0.05 mJy 	& 5.26$ \pm $0.57 mJy	\\ \hline
$20^h23^m02^s$ 	&               	&                	\\
38\degr42\arcmin27\arcsec 	& 5.17$ \pm $0.05 mJy 	& 6.3$ \pm $0.6 mJy	\\ \hline
$20^h21^m38^s$ 	&               	&               	\\
38\degr47\arcmin18\arcsec 	& 2.22 $\pm$ 0.06 mJy 	& 1.7 $\pm$ 0.5 mJy	\\ \hline
\end{tabular}
\caption{A comparison between GMRT 1160-MHz and VLA 1465-MHz flux densities is
  shown in this table for the four brightest sources around G76.9+1.0,
  excluding 4C39.61. The VLA 1465 MHz C/D array data was taken on 17
  May 1988. The leftmost column gives the right ascension and
  declination in J2000 coordinates.} 
\label{vlafluxcompare}
\end{table}

\section{\emph{Chandra} X-ray data and analysis}
\label{xray}
\begin{figure}   
\begin{center}
    \includegraphics[scale=0.47]{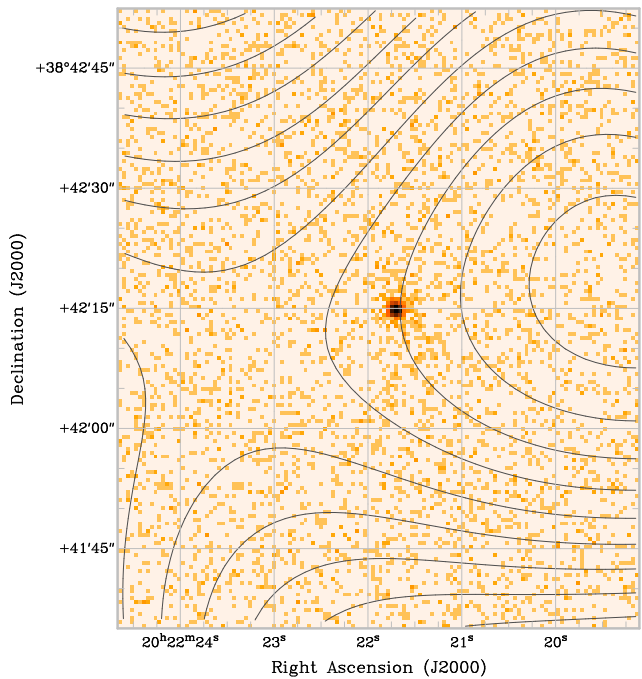}     
    \caption{The point source seen at the same position in the raw
      $0.3-7{\rm~keV}$ \emph{Chandra} image,
%. The precise location of
 %     the point is $\alpha: 20^{h}22^{m}21.692^{s},\ \delta:
  %    38\degr42\arcmin14.81\arcsec$(J2000), 
     establishing unambiguously that the radio
      counterparts at 618~MHz and 1160~MHz are the same object. The
      GMRT 618 MHz contours are overlaid to highlight the location of
      the compact X-ray source with respect to the diffuse radio
      bridge of emission.} 
\label{chimage}
\end{center}
\end{figure}

\begin{figure}
   \includegraphics[width=8.5cm]{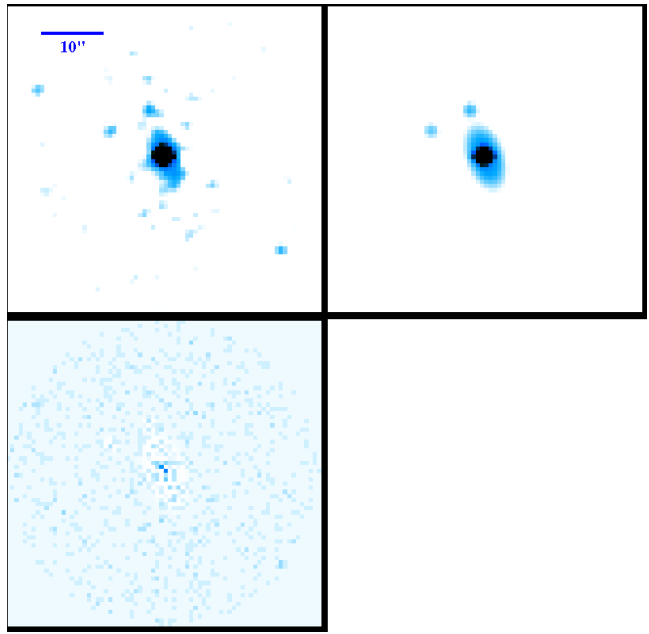}
   \caption{The observed ACIS-S image ({\it top left panel}), the model image
  ({\it top right panel}) in the $0.3-7{\rm~keV}$ band. The residual image
  (observed -- model) is shown in the bottom panel.} 
\label{chandra_image}
\end{figure}

ACIS-S detector data from the 01 Aug 2005 \emph{Chandra} observation 
of G79.6+1.0 (exposure time $54.4{\rm~ks}$) was downloaded from the 
archive. The unprocessed \emph{Chandra} X-ray image in the $0.3-7{\rm~keV}$ 
band is shown in Fig.~\ref{chimage}. An unresolved source is detected
at $\alpha: 20^{h}22^{m}21.7^{s},\ \delta:
38\degr42\arcmin14.8\arcsec, J2000$, establishing unambiguously that
the radio counterparts at 618~MHz and 1160~MHz are the same
object. The overall 90\% uncertainty circle of Chandra X-ray absolute position has a
radius of $0.6\arcsec$. 
We reprocessed the Level-1
event files from the observation using the latest version of the
\emph{Chandra} Interactive Analysis of Observations software (CIAO
4.2) and calibration database (CALDB 4.2), and filtered them using the good
time interval files to obtain the Level-2 event files. We extracted
the ACIS-S image in the $0.3-7{\rm~keV}$ band and smoothed by
convolving with a Gaussian of FWHM$=2{\rm~pixels}$. The smoothed image
is shown in Fig.~\ref{chandra_image}. A point source coincident 
with the unresolved radio source is seen. In addition, the ACIS-S image 
also showed weak extended emission around the point source. 

\begin{figure}
    \includegraphics[scale=0.33,angle=-90]{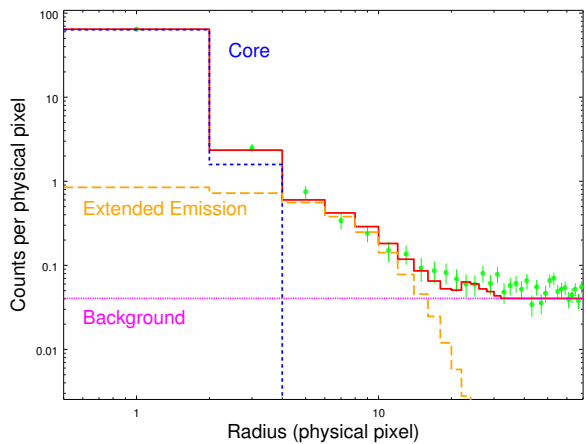}
\caption{Radial profiles of the central core and the nebular emission
  based on the spatial modeling of the \emph{Chandra} $0.3-7{\rm~keV}$
  image.} 
\label{radprof}
\end{figure}

The extended emission can be identified as a synchrotron nebula around 
the neutron star. In order to find the relative contribution of the 
unresolved X-ray source and the surrounding nebula, we 
carried out spatial modeling of the observed ACIS-S image using 
CIAO's modeling and fitting package, Sherpa 4.2. We modelled the 
unresolved core as a narrow, two-dimensional Gaussian and the 
nebula with a broad, two-dimensional Gaussian. We also 
used a constant component to account for the background. Additionally, 
we used two narrow Gaussians to account for two nearby point sources 
seen in Fig.~\ref{chandra_image}. We carried out the fitting using the
C-statistics appropriate for low-count data. The best-fitting model
resulted in a strong unresolved core and an elongated extended
emission with ellipticity of $\epsilon=0.48 \pm 0.06$ and the
ellipticity angle $\theta=108.3 \pm 4.6{\rm~deg}$; all errors are
$1\sigma$.  The best-fitting model image ({\it top right panel}) and
the residual image ({\it bottom left panel}) are also shown in
Fig.~\ref{chandra_image}. The extended emission contributes only $15\%$
to the total X-ray emission from G79.6+1.0 in the $0.3-7{\rm~keV}$
band. We have also plotted the radial profiles
of the central core and the extended emission in
Fig.~\ref{radprof}. Clearly, the X-ray nebula extends upto
$\sim7\arcsec$, while the core is confined to within $2\arcsec$. 

\begin{figure}
  \includegraphics[width=6.75cm,angle=-90]{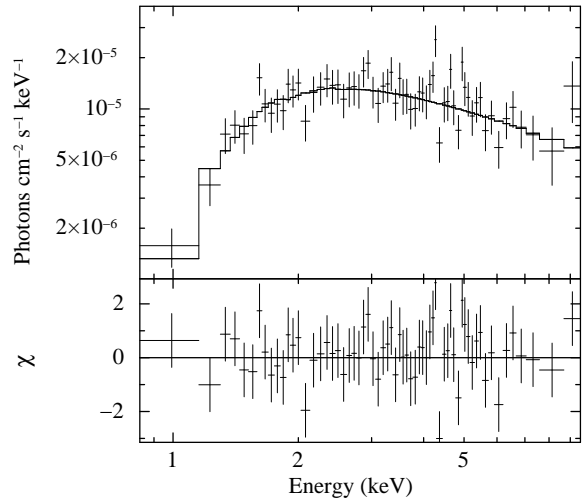}
  \caption{The observed ACIS-S spectral data, the best-fitting model
    consisting of an absorbed powerlaw component, and the deviations
    of the observed data from the model.} 
  \label{chandra_spec}
\end{figure}

The source spectrum was extracted from the Level-2 events using a
circular region of radius 20 pixels = $9.84\arcsec$. We also extracted
a background spectrum from four circles of radii $20{\rm~pixels}$ in
the nearby source-free regions. Instrumental responses were generated
(the redistribution matrix and the ancillary response files) using the
tasks \emph{mkacisrmf} and \emph{mkarf}. The spectral data were
grouped to a minimum counts of 20 per bin and spectral analysis
performed with the XSPEC (version 12.6) package, using the $\chi^2$
statistics. The source spectral extraction region includes both the
unresolved core and the extended synchrotron nebula. For an energetic
pulsar like G76.9+1.0 with extedned emission, X-ray emission should be
dominated by magnetospheric emission. Therefore, we used a simple
power-law model, modified by the Galactic absorption, to describe the X-ray
emission from G79.6+1.0. The best-fitting model resulted in the
absorption column $N_H= (1.5 \pm 0.3)\times10^{22}{\rm~cm^{-2}}$ and
power-law photon index $\Gamma=0.9\pm0.2$ with minimum $\chi^2=62.5$
for 61 degrees of freedom. Here the errors are quoted at the $90\%$
confidence level.  The unfolded \emph{Chandra} spectrum, the best
fit spectral model and the deviations of the observed data from the
model are shown in Fig.~\ref{chandra_spec}. The observed 
$0.5-8{\rm~keV}$ flux is $4.6\times10^{-13}{\rm~ergs~cm^{-2}~s^{-1}}$
and the corresponding unabsorbed flux is $5.9\times10^{-13}$. Addition
of a blackbody component does not improve the fit ($\Delta\chi^2=-0.7$
for two additional parameters). Similarly, an additional power-law
component is not required by the data.

\section{The nature of the central point source}
\label{summary}
One can compute the approximate age of the remnant and therefore of
the neutron star. We invoke equipartition between the magnetic field
and the particles to obtain the equipartition pressure. An unbroken
power-law with $\alpha = -0.61$ in the range 10~MHz-10~GHz gives a
radio synchrotron luminosity of $L_r \sim 8 \times 10^{36}
erg\ s^{-1}$. The pressure obtained from the equipartition magnetic
field - $B_{eq} \sim 0.6\ mG$ - is P $\sim 1.5 \times 10^{-8}
erg\ cm^{-3}$. The mean expansion velocity, $\sim 1000\ km/s$,
gives an estimate of the age of the neutron star as $\sim 5
\times \sqrt{n_p}\ kyr$, where $n_p$ is the particle
number density. For a mean particle density of $1\ cm^{-3}$, the approximate age of the remnant is 5 kyr.

\subsection{Non-detectability of the radio pulsar}
\label{ssec:nopulse}
The location, steep radio spectrum as well as the X-ray counterpart
of the the point source make a strong case for assuming that it arises
from the central pulsar associated with this SNR. The lack of detection
of pulsed emission is however puzzling. \citet*{lorimer1998} had also
failed to detect pulsed emission from G76.9+1.0, however their limits
are substantially weaker than those presented here. In
Sec.~\ref{gmrt1160} we had advanced three possibilities for the 
lack of pulsed emission. Out of the three, the most likely cause for
the lack of pulsed emission appears to be temporal broadening. The 
expected temporal broadening using equation~(7) from \citet{bhat2004}
is

\begin{equation}
\log(\tau_d) \approx a + b\ \log({\rm DM}) + c\ (\log({\rm DM}))^2 -
\alpha\ \log(\nu)
\end{equation}

where $a = -6.46, b = 0.154, c = 1.07$ and $\alpha$ = 3.86. Here, the 
temporal broadening $\tau_d$ is in ms and observing frequency $\nu$ is 
in GHz. For the DM range of $250\ \!-\ \!500\ pc\ \!cm^{-3}$ the expected pulse broadening
at 618~MHz and 1170~MHz are $8-380\ \!$ms and $0.6-30\ \!$ms
respectively. Clearly 
if the DM lies in the upper end of the likely range and the pulsar 
has a period in the $10-30\ \!$ms range, the pulse broadening will be
comparable to the pulse period, making it very difficult to detect
the pulsar. We note in this context that during the pulsar search
candidates were identified by summing up to 16 harmonics, and hence
even if only the fundamental alone was statistically significant
the pulsar should have shown up as a candidate in our
search. Accounting temporal broadening alone as being responsible for
non-detection of the pulse, we considered sensitive higher frequency follow-up
observations, perhaps with the GBT. However, going through the GBT
data archives we learnt that this object had already been observed and
a pulsar has been detected. The period of the pulsar(J2022+3842) is
$\sim24$ms, and the DM is $430\ pc\ \!cm^{-3}$(Z.~Arzoumanian and S.~M.~Ransom,
private communication), consistent with our non-detections at lower
frequencies. Their 2-GHz flux of $\sim 75\ \mu$Jy agrees well with the
spectral index measured with the GMRT. This establishes beyond any
ambiguity that the pulsar non-detections at the GMRT were not
sensitivity-limited, but because of the large DM and the lower
observing frequencies, the pulsed signal would be appreciably
broadened. It seems a reasonable conclusion, therefore, that the radio
emission seen in the GMRT images as the unresolved source comes from
the pulsar, but that only because of temporal broadening we do not
detect any pulsed emission. We turn now to the discussion of the
diffuse emission seen in the X-ray image.    

\subsection{The geometry of the wind nebula}
The X-ray image in Fig.~\ref{chandra_image} shows diffuse emission, whose
major axis is approximately aligned to the bipolar structure seen
in the radio images. No radio counterpart to this diffuse X-ray emission
is obvious in our data: this may partly be due to the difficulties in
disentangling it from the more extended emission in the SNR.

\begin{figure}
\begin{center}
\includegraphics[scale=0.6]{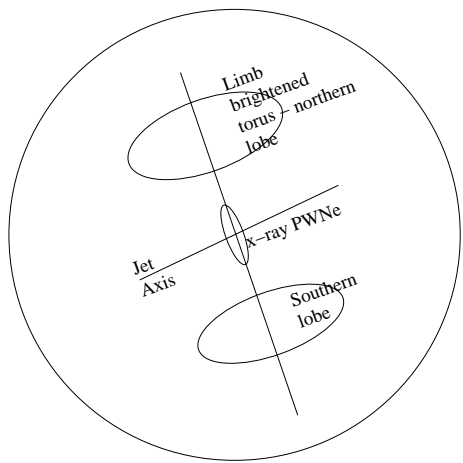}
\caption{The geometry of the x-ray wind nebula seen together with the
  radio hotspots within the supernova remnant G79.6+1.0. The lobes
  represent sites where charged particles radiate in radio
  synchrotron. The torus is viewed at an angle almost normal to its
  axis along which the jet roughly aligns. The radio hotspots arise
  from ``limb brightening'' attributed to the viewing angle.}
\label{snrmodel}
\end{center}
\end{figure}

The pulsar is located at the center of the diffuse
X-ray emission, but offset from the central depression in the radio
image. In the case of the PWNe DA495 there is a similar offset between
the location of the pulsar and the central depression in the bipolar
emission, although the alignment with the diffuse X-ray emission is 
not clear \citep{kothes2008,zaven2008}. \citet{kothes2008} interpret
the maxima in the radio bipolar structure as lying along the rotation axis.
They also suggest that the minima in the radio image correspond to 
dense gas emitted earlier by an equatorial wind of the parent star. 
Since synchrotron particles cannot penetrate this dense material 
this leads to a minimum in the radio emission. Guided
by the ellipticity of the diffuse component of the X-ray emission in G76.9+1.0 we suggest
a different interpretation for this PWNe. We assume that the
diffuse elliptical component of the 
X-ray emission traces the equatorially-flowing wind from the pulsar, and this
wind continues to produce a radio synchrotron torus at large radii, which we see
as a ``limb brightened'' bipolar structure (see Fig.~\ref{snrmodel}). 
The minimum in the radio emission would then be perpendicular to 
the equatorial wind, i.e. along the rotation axis of the pulsar as
projected on the sky.
In this context it is interesting to note that in the radio image 
there is a smaller secondary minimum diametrically opposite the 
central one (i.e. at $\alpha 
\sim 20^h22^m35^s, \delta \sim 38^o42^{'}$), similar to the situation
in DA495. In our model this would arise because in the equatorial
wind one would expect a minimum in the emission along the rotation
axis(or its projection on the sky). Interestingly, in the X-ray image
(Fig.~\ref{chandra_image}) there is a jet-like feature perpendicular
to the diffuse emission, i.e. aligned with the minima which we posit
to lie along the rotation axis. Narrow, jet-like features in PWNe are
generally posited to lie along the rotation axis of the pulsar
\citep*{lai2001}. It has also been suggested that the feature lies
along the projection of the magnetic axis on to the rotation axis
\citep{rad2001}, in which case particle beams emanating from the
magetic poles, offset from the rotation axis, trace out a pair of
arcs, as seen in the Vela, and the wind eventually settles down on the
equatorial plane at large radii. In either case one would expect the
observed perpendicularity of the jet and the equatorial plane. In our
picture, the polarization seen in G76.9+1.0  by \citet{landecker1993}
would be caused by the magnetic field becoming approximately radial at
large distances from the pulsar.  

\section{Conclusions}
We report the serendipitous discovery of an unresolved, steep spectrum
radio source in the SNR G76.9+1.0. The mean measured radio position of this point
  source is $\alpha =
  20^{h}22^{m}21.69^{s}\ \pm\ 0.11^{s},\ \delta =
  38\degr42\arcmin14.8\arcsec\ \pm\ 1.7\arcsec, J2000$. Analysis of
  archival {\it Chandra} X-ray data shows this to be coincident with
  an unresolved X-ray source. Despite a deep search no pulsed emission
  was detected at 610~MHz and 1160~MHz with the GMRT. However the
  pulsar has been seen with the GBT at 2~GHz. This has been understood
  as the consequence of the temporal broadening of the pulse due to
  the large dispersion measure along the line of sight to the pulsar
  and the low radio frequencies with which it was observed at the
  GMRT. The X-ray emission also shows a diffuse elliptical structure
  aligned along the bipolar structure seen in the radio. We suggest
  that these structures arise because of an equatorial wind from the
  pulsar. We underline the usefulness of a high-resolution radio
  imaging study in locating and prospecting for pulsar candidates in
  supernova remnants, which could otherwise be missed in a time-series
  pulsar search observation. 

\section{Acknowledgements}
We thank the GMRT staff for having made possible 
the observations used for the research reported in this paper. The
GMRT is run by the National  Centre for Radio Astrophysics of the Tata
Institute of Fundamental Research.
We are grateful to Z.~Arzoumanian and S.~Ransom for communicating
the results of their GBT pulsar search prior to publication. We thank
the referee for pointing us to \citet{zaven2010}.
VRM thanks various colleagues for several fruitful discussions during
the course of this research. He also thanks the Raman Research
Institute, Bangalore, India for their hospitality while writing part
of this paper. Stimulating conversations on this project with
A.~A.~Deshpande are also gratefully acknowledged.

\end{document}